\documentclass[11pt,letterpaper]{amsart}
\usepackage{amsmath,amssymb,amsthm,mathscinet}
\usepackage{enumerate,multirow,color,colortbl,tikz}
\renewcommand{\epsilon}{\varepsilon}
\renewcommand{\phi}{\varphi}
\DeclareMathOperator{\Tr}{Tr}
\DeclareMathOperator{\PC}{PC}
\DeclareMathOperator{\ADF}{ADF}
\DeclareMathOperator{\AMF}{AMF}
\DeclareMathOperator{\CDF}{CDF}
\DeclareMathOperator{\CMF}{CMF}
\DeclareMathOperator{\PSC}{PSC}
\newcommand{\C}{{\mathbb C}}
\newcommand{\F}{{\mathbb F}}
\newcommand{\Z}{{\mathbb Z}}
\newcommand{\unitsymbol}{*}
\newcommand{\Cu}{\C^{\unitsymbol}}
\newcommand{\Fp}{\F_p}
\newcommand{\Fpu}{\F_p^{\unitsymbol}}
\newcommand{\Fpum}{\F_p^{\unitsymbol m}}
\newcommand{\Fputwo}{\F_p^{\unitsymbol 2}}
\newcommand{\Fpufour}{\F_p^{\unitsymbol 4}}
\newcommand{\cycquot}{\Fpu/\Fpum}
\newcommand{\Fpn}{\F_{p^n}}
\newcommand{\Fq}{\F_q}
\newcommand{\conj}[1]{\overline{#1}}
\newcommand{\sums}[1]{\sum_{\substack{#1}}}
\newcommand{\norm}[2]{\Vert{#1}\Vert_{#2}}
\newcommand{\normp}[3]{\norm{#1}{#2}^{#3}}
\newcommand{\normtt}[1]{\normp{#1}{2}{2}}
\newcommand{\normtf}[1]{\normp{#1}{2}{4}}
\newcommand{\normff}[1]{\normp{#1}{4}{4}}
\newcommand{\cnrd}[2]{\frac{\normtt{{#1} {#2}}}{\normtt{#1}  \normtt{#2}}}
\newcommand{\finishtablerow}{\\ \noalign{\vskip-0.3pt}}
\definecolor{tablecolor}{rgb}{0.85,0.85,0.85}
\newcommand{\colortablerow}{\rowcolor{tablecolor}[5.1pt]}
\newtheorem{question}{Question}
\title{Sequences with Low Correlation}
\author{Daniel J. Katz}
\address{Department of Mathematics, California State University, Northridge, \: United States}\thanks{This paper is based on work supported in part by the National Science Foundation under Grant DMS 1500856.}
\date{12 June 2018}
\begin{document}
\begin{abstract}
Pseudorandom sequences are used extensively in communications and remote sensing.
Correlation provides one measure of pseudorandomness, and low correlation is an important factor determining the performance of digital sequences in applications.
We consider the problem of constructing pairs $(f,g)$ of sequences such that both $f$ and $g$ have low mean square autocorrelation and $f$ and $g$ have low mean square mutual crosscorrelation.
We focus on aperiodic correlation of binary sequences, and review recent contributions along with some historical context.
\end{abstract}
\maketitle
\section{Introduction}
Sequences with low correlation play many roles in technology, including remote sensing, design of scientific instruments, operation of communications networks,
and acoustic design.
The monographs of Golomb, Gong, and Schroeder \cite{Golomb,Golomb-Gong,Schroeder} give some sense of the broad sweep of their applications.  Golomb \cite[p.~25]{Golomb} used correlation as a measure of pseudorandomness, a concept of significance for cryptography that has been developed extensively by Mauduit and S\'arkozy in \cite{Mauduit-Sarkozy} and further works.
Here we give an overview of recent progress on the problem of designing binary sequence pairs where both sequences have low aperiodic autocorrelation and the two sequences of the pair have low mutual aperiodic crosscorrelation.

This paper is organized as follows: Sections \ref{Katherine}, \ref{Melissa}, and \ref{Lisa} give the basic definitions (of sequences, correlation, and merit factors).
Section \ref{Gary} lists some constructions of sequence families with low mean square autocorrelation.
Sections \ref{Theresa}, \ref{Ursula}, and \ref{Lawrence} describe how the constructions are done, and provide more details about autocorrelation performance.
Section \ref{Rebecca} examines the question of low mean square crosscorrelation, and Section \ref{Nina} discusses a combined measure (called the Pursley-Sarwate criterion) of autocorrelation and crosscorrelation performance of a sequence pair.
Section \ref{Orestes} discusses families of sequence pairs with low Pursley-Sarwate criterion, and Section \ref{Konrad} concludes with open questions.

\section{Sequences}\label{Katherine}

If $\ell$ and $m$ are positive integers, an {\it additive $m$-ary sequence of length $\ell$} is an $\ell$-tuple of elements of the additive group of $\Z/m\Z$, that is,
\begin{equation}\label{Alice}
a=(a_0,a_1,\ldots,a_{\ell-1}) \in (\Z/m\Z)^\ell.
\end{equation}
When $m=2$ we have an {\it additive binary sequence}, that is, an element of $\F_2^\ell$.

A {\it multiplicative $m$-ary sequence of length $\ell$} is an $\ell$-tuple of $m$th roots of unity in $\C$, that is,
\begin{equation}\label{Bob}
b=(b_0,b_1,\ldots,b_{\ell-1}) \in \mu_m^\ell,
\end{equation}
where $\mu_m=\{e^{2\pi i j/m}: 0 \leq j < m\}$ is the multiplicative group of $m$th roots of unity in $\C$.
Most often we have $m=2$, so $\mu_2=\{1,-1\}$; this gives {\it multiplicative binary sequences}, which we shall just call {\it binary sequences}.

Consider the group homomorphism $\epsilon\colon \Z/m\Z \to \mu_m$ with $\epsilon(x)=e^{2\pi i x/m}$.
If the sequences $a$ and $b$ of \eqref{Alice} and \eqref{Bob} are related by $b_k=\phi(a_k)$ for every $k$, then we say that {\it $b$ is the multiplicative version of $a$}, and equivalently, that {\it $a$ is the additive version of $b$}.

For the purposes of this paper, it will be more convenient to consider sequences in their multiplicative guise.
Furthermore, we shall identify the sequence $f=(f_0,\ldots,f_{\ell-1}) \in \C^\ell$ in multiplicative form with the polynomial $f(z)=f_0+f_1 z+ \cdots + f_{\ell-1} z^{\ell-1} \in \C[z]$, whose coefficients are the terms of the sequence $f$.
This identification makes calculations easier and we shall see in Section \ref{Lisa} that it forms a bridge between the study of correlation and harmonic analysis that has proved fruitful in these studies.

\section{Correlation}\label{Melissa}
Correlation is a measure of the similarity between the various shifted versions of a pair of sequences.
When the sequences of the pair are the same, we are comparing a sequence to shifted versions of itself, which is self-correlation, or autocorrelation.
Truly random sequences should have low correlation with shifted versions of themselves (unless the shift is zero) and of each other, so we demand that our pseudorandom sequences also have low correlation.

Let us now define correlation precisely.
For two sequences
\begin{align}\begin{split}\label{Denise}
f & = (f_0,f_1,\ldots,f_{\ell-1}) \in \C^\ell \\
g & = (g_0,g_1,\ldots,g_{\ell-1}) \in \C^\ell,
\end{split}
\end{align}
and $s\in \Z$, the {\it aperiodic crosscorrelation of $f$ with $g$ at shift $s$}, denoted $C_{f,g}(s)$, is defined by
\[
C_{f,g}(s) = \sum_{j \in \Z} f_{j+s} \conj{g_j},
\]
where we use the convention that $f_j=g_j=0$ when $j \not\in \{0,1,\ldots,\ell-1\}$, so that the above sum only has a finite number of nonzero entries.
We index over $\Z$ because the underlying translation operation involved in aperiodic crosscorrelation is a non-cyclic shift.
We can view the aperiodic correlation of $f$ with $g$ at shift $s$ as the inner product between the overlapping portions of $f$ and $g$ when $g$ is shifted $s$ places to the right relative to $f$, as shown in Figure \ref{Caesar}.
\begin{center}
\begin{figure}[!ht]
{\footnotesize\begin{tikzpicture}[scale=1.05]
\draw (-4,0) rectangle (-3,1);
\node at (-3.5,0.5) {$f_0$};
\draw (-3,0) rectangle (-2,1);
\node at (-2.5,0.5) {$f_1$};
\node at (-1.5,0.5) {$\cdots$};
\draw (-1,0) rectangle (0,1);
\node at (-0.5,0.5) {$f_{s-1}$};
\draw (0,0) rectangle (1,1);
\node at (0.5,0.5) {$f_s$};
\draw (1,0) rectangle (2,1);
\node at (1.5,0.5) {$f_{s+1}$};
\node at (2.5,0.5) {$\cdots$};
\draw (3,0) rectangle (4,1);
\node at (3.5,0.5) {$f_{\ell-1}$};
\draw (0,-1.5) rectangle (1,-0.5);
\node at (0.5,-1) {$g_0$};
\draw (1,-1.5) rectangle (2,-0.5);
\node at (1.5,-1) {$g_1$};
\node at (2.5,-1) {$\cdots$};
\draw (3,-1.5) rectangle (4,-0.5);
\node at (3.5,-1) {$g_{\ell-s-1}$};
\draw (4,-1.5) rectangle (5,-0.5);
\node at (4.5,-1) {$g_{\ell-s}$};
\node at (5.5,-1) {$\cdots$};
\draw (6,-1.5) rectangle (7,-0.5);
\node at (6.5,-1) {$g_{\ell-2}$};
\draw (7,-1.5) rectangle (8,-0.5);
\node at (7.5,-1) {$g_{\ell-1}$};
\end{tikzpicture}}
\caption{Aperiodic correlation of $f$ with $g$ at shift $s > 0$}\label{Caesar}
\end{figure}
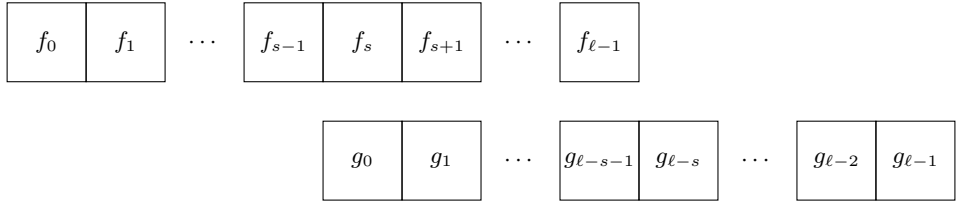
\end{center}
Let us identify $f$ and $g$ of \eqref{Denise} with the polynomials $f(z)=f_0+f_1 z+\cdots + f_{\ell-1} z^{\ell-1}$ and $g(z)=g_0+g_1 z + \cdots +g_{\ell-1} z^{\ell-1}$, respectively, as discussed at the end of Section \ref{Katherine}.
Furthermore, let us adopt the convention that for any Laurent polynomial
\[
a(z)=\sum_{j \in \Z} a_j z^j,
\]
in the ring $\C[z,z^{-1}]$ of Laurent polynomials over $\C$, the {\it conjugate of $a(z)$} is defined to be
\begin{equation}\label{Nancy}
\conj{a(z)}=\sum_{j \in \Z} \conj{a_j} z^{-j},
\end{equation}
where $\conj{a_j}$ is the usual complex conjugate of $a_j$.
Then it is not difficult to show that
\[
f(z) \conj{g(z)} = \sum_{s \in \Z} C_{f,g}(s) z^s,
\]
that is, the crosscorrelation of $f$ with $g$ at shift $s$ is the coefficient of $z^s$ in the product $f(z) \conj{g(z)}$.
This interpretation allows us to discover quite easily the following basic symmetry of aperiodic correlation:
\begin{equation}\label{Joan}
C_{f,g}(s) = \conj{C_{g,f}(-s)}.
\end{equation}   
    
There is also a periodic version of correlation that treats our sequences \eqref{Denise} as periodically repeating every $\ell$ terms.
In this case, $\Z/\ell\Z$ is the natural set for indexing sequence terms and expressing shifts, reflecting the cyclic nature of the sequences and the shifting.
Then for any $s \in \Z/\ell\Z$, the {\it periodic crosscorrelation of $f$ with $g$ at shift $s$}, denoted $\PC_{f,g}(s)$, is defined by
\[
\PC_{f,g}(s) = \sum_{j \in \Z/\ell\Z} f_{j+s} \conj{g_j}.
\]
We can view the periodic correlation of $f$ with $g$ at shift $s$ as the inner product of $f$ with $g$ when $g$ is cyclically shifted $s$ places to the right relative to $f$, as shown in Figure \ref{Elizabeth}.
\begin{center}
\begin{figure}[!ht]
{\footnotesize\begin{tikzpicture}[scale=1.05]
\draw (-4,0) rectangle (-3,1);
\node at (-3.5,0.5) {$f_0$};
\draw (-3,0) rectangle (-2,1);
\node at (-2.5,0.5) {$f_1$};
\node at (-1.5,0.5) {$\cdots$};
\draw (-1,0) rectangle (0,1);
\node at (-0.5,0.5) {$f_{s-1}$};
\draw (0,0) rectangle (1,1);
\node at (0.5,0.5) {$f_s$};
\draw (1,0) rectangle (2,1);
\node at (1.5,0.5) {$f_{s+1}$};
\node at (2.5,0.5) {$\cdots$};
\draw (3,0) rectangle (4,1);
\node at (3.5,0.5) {$f_{\ell-1}$};
\draw (0,-1.5) rectangle (1,-0.5);
\node at (0.5,-1) {$g_0$};
\draw (1,-1.5) rectangle (2,-0.5);
\node at (1.5,-1) {$g_1$};
\node at (2.5,-1) {$\cdots$};
\draw (3,-1.5) rectangle (4,-0.5);
\node at (3.5,-1) {$g_{\ell-s-1}$};
\draw (-4,-1.5) rectangle (-3,-0.5);
\node at (-3.5,-1) {$g_{\ell-s}$};
\draw (-3,-1.5) rectangle (-2,-0.5);
\node at (-2.5,-1) {$g_{\ell-s+1}$};
\node at (-1.5,-1) {$\cdots$};
\draw (-1,-1.5) rectangle (0,-0.5);
\node at (-0.5,-1) {$g_{\ell-1}$};
\end{tikzpicture}}
\caption{Periodic correlation of $f$ with $g$ at shift $s$}\label{Elizabeth}
\end{figure}
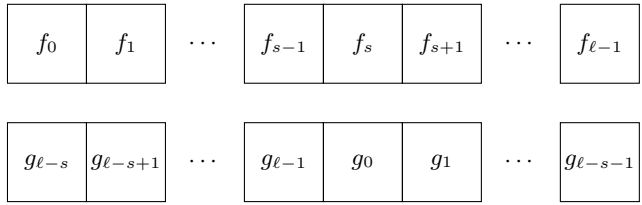
\end{center}
If $s \in \Z$, then we interpret $\PC_{f,g}(s)$ as $\PC_{f,g}(\sigma)$ where $\sigma \in \Z/\ell\Z$ is the congruence class of $s$ modulo $\ell$.
In this case one can see that for any $s \in \Z$, we have
\[
\PC_{f,g}(s) = \sums{t \in \Z \\ t \equiv s \!\!\! \pmod{\ell}} C_{f,g}(t),
\]
where at most two of these terms can be nonzero, and in particular, if $0 \leq s < \ell$, then
\begin{equation}\label{Francis}
\PC_{f,g}(s) = C_{f,g}(s)+C_{f,g}(s-\ell).
\end{equation}
One can obtain a polynomial interpretation of periodic crosscorrelation by regarding our sequences as lying in the ring $\C[z]/(z^\ell-1)$ rather than $\C[z]$.
The notion of a conjugate of a Laurent polynomial from \eqref{Nancy} carries over naturally to $\C[z]/(z^\ell-1)$: negative powers of $z$ can be reinterpreted as positive powers since $z^\ell=1$ in this ring, and the ideal $(z^\ell-1)$ is closed under our conjugation since $\conj{z^\ell-1}=-z^{-\ell}(z^\ell-1)$.
Then one can show that
\[
f(z) \conj{g(z)} \equiv \sum_{s \in \Z/\ell\Z} \PC_{f,g}(s) z^s \pmod{z^\ell-1},
\]
and from this prove the symmetry
\[
\PC_{f,g}(s) = \conj{\PC_{g,f}(-s)}.
\]

Periodic correlation is more mathematically tractable than aperiodic correlation.
For example, when we consider sequences derived from finite field characters (see Sections \ref{Theresa} and \ref{Ursula}), the periodic correlation values are complete character sums, while the aperiodic correlation values are incomplete character sums, which are much more difficult to handle.
Equation \eqref{Francis} shows that the magnitude of any periodic correlation cannot be more than twice as large as the largest magnitude of any aperiodic correlation value.
In consequence of this, Boehmer \cite[p.~157]{Boehmer} points out that having low periodic correlation at all shifts is a necessary but not sufficient condition for having low aperiodic correlation at all shifts.
She then enunciates a design technique that has been used widely in attempts to design sequences with low aperiodic correlation: design sequences with low periodic correlation and hope that some of these will also have low aperiodic correlation.

Earlier we had mentioned autocorrelation, or correlation of a sequence with itself.
If $f$ is the sequence in \eqref{Denise} and $s \in \Z$, then the {\it aperiodic autocorrelation of $f$ at shift $s$} is just the aperiodic crosscorrelation of $f$ with itself at shift $s$, that is,
\[
C_{f,f}(s)=\sum_{j \in \Z} f_{j+s} \conj{f_j}.
\]
And for $s \in \Z/\ell\Z$, the {\it periodic autocorrelation of $f$ at shift $s$} is just the periodic crosscorrelation of $f$ with itself at shift $s$, that is,
\[
\PC_{f,f}(s)=\sum_{j\in \Z/\ell\Z} f_{j+s} \conj{f_j}.
\]
Note that if the shift is zero in either the aperiodic or periodic case, then 
\begin{equation}\label{Imogene}
C_{f,f}(0)=\PC_{f,f}(0) = \sum_{j=0}^{\ell-1} |f_j|^2,
\end{equation}
which is the squared Euclidean norm of $f$ if it is regarded as a vector in $\C^\ell$.
If all the terms of $f$ are of unit magnitude, we say that $f$ is a {\it unimodular sequence}.
For example, all $m$-ary sequences are unimodular, since their terms are roots of unity.
If $f$ is unimodular, then
\[
C_{f,f}(0)=\PC_{f,f}(0) = \ell,
\]
which is the length of the sequence, and is naturally as large as a correlation value involving unimodular sequences of length $\ell$ could possibly be.
On the other hand, what we know about random walks suggests that typical correlation values for randomly selected binary sequences of length $\ell$ (with uniform probability distribution) should not have magnitudes much larger than $\sqrt{\ell}$.

\section{Demerit factors and merit factors}\label{Lisa}

In a multi-user communications network, one can modulate the messages of the various users with different signature sequences.
For efficient operation, it is desirable that the family of signature sequences used should have the following properties:
\begin{enumerate}[(i)]
\item\label{George} Each sequence $f$ should have low magnitude autocorrelation $|C_{f,f}(s)|$ at all nonzero shifts (all $s\not=0$).
\item\label{Hilary} Each pair $(f,g)$ of sequences should have low magnitude crosscorrelation $|C_{f,g}(s)|$ at all shifts $s$.
\end{enumerate}
Notice that it is the magnitude of the correlation value that is typically considered, since often the argument is not discernible in our systems.
In view of our comments on random sequences at the conclusion of the previous section, a typical correlation value can be considered small if it does not have magnitude much larger than the square root of the length of the sequences involved.
Condition \eqref{George} helps the communications system maintain synchronization with the user represented by sequence $f$: the sharp difference between correlation of at shift $0$ (when the sequence is aligned with a reference copy of itself) and at nonzero shifts (when it is not aligned) allows one to obtain very accurate timings.
Condition \eqref{Hilary} prevents the output from the user represented by sequence $f$ from being confused with any output from the user represented by sequence $g$, regardless of any delays between these two signals.

In this paper, we consider the simplest possible case of this design problem for sequences with low autocorrelation and crosscorrelation, that is, we ask for pairs of sequences such that $|C_{f,f}(s)|$ and $|C_{g,g}(s)|$ are low for all $s\not=0$ and $|C_{f,g}(s)|$ is low for all $s$.
We could rate overall smallness of correlation in various ways.
One method is to rate crosscorrelation performance for a sequence pair $(f,g)$ by the worst case: the {\it peak crosscorrelation of $f$ with $g$} is
\[
\max_{s \in \Z} |C_{f,g}(s)|,
\]
and we would want this to be small.  

After studying this and some other common measures of smallness of crosscorrelation, K\"arkk\"ainen \cite[p.~149]{Karkkainen} expresses the view that one gets a better notion of likely performance from a mean square measure.
Accordingly, for sequence pair $(f,g)$, we define the {\it crosscorrelation demerit factor of $f$ and $g$} as
\[
\CDF(f,g) = \frac{\sum_{s\in\Z} |C_{f,g}(s)|^2}{|C_{f,f}(0)| \cdot |C_{g,g}(0)|},
\]
which, in view of \eqref{Imogene}, is the sum of squared magnitudes of crosscorrelation values for the sequence pair we obtain from $f$ and $g$ if we scale each of them to have unit Euclidean magnitude.
We should note that $\CDF(f,g)=\CDF(g,f)$ because of \eqref{Joan}.
Normally $f$ and $g$ are unimodular sequences of the same length $\ell$, so the denominator of the $\CDF$ is simply $\ell^2$.
Since we want every term in the numerator to be as small as possible, a large $\CDF$ indicates poor performance.
If one wants a measure that is larger for good sequence pairs, one defines the {\it crosscorrelation merit factor of $f$ and $g$} to be
\[
\CMF(f,g)=\frac{1}{\CDF(f,g)}.
\]
We have analogous measures for autocorrelation.
If $f$ is a sequence, then the {\it autocorrelation demerit factor of $f$} is defined to be
\begin{equation}\label{Paul}
\ADF(f) = \frac{\sums{s \in \Z \\ s\not=0} |C_{f,f}(s)|^2}{|C_{f,f}(0)|^2} = \CDF(f,f)-1,
\end{equation}
where one should note that we omit the autocorrelation at shift $0$ in the numerator.
This is because $C_{f,f}(0)$ is always large, and we want it to be large, so that it should not be construed as contributing to the demerit factor.
And the {\it autocorrelation merit factor of $f$} is just the reciprocal of the demerit factor,
\[
\AMF(f) = \frac{1}{\ADF(f)}.
\]
Naturally we want to make the autocorrelation demerit factor small, or equivalently, to make the autocorrelation merit factor large.

Recall from the end of Section \ref{Katherine} that we always identify the sequence $f=(f_0,\ldots,f_{\ell-1}) \in \C^\ell$ with the polynomial $f(z)=f_0+f_1 z+ \cdots + f_{\ell-1} z^{\ell-1} \in \C[z]$.
We shall now see how this point of view relates to merit factors.
For any real number $r \geq 1$ and any function $f$ defined on the complex unit circle, we define the {\it $L^r$ norm of $f$ on the complex unit circle} to be
\[
\norm{f}{r} = \left(\frac{1}{2\pi}\int_0^{2\pi} |f(e^{i\theta})|^r d\theta\right)^{1/r},
\]
provided that this integral exists (as it certainly will when $f$ is a Laurent polynomial).

Then one can show that the crosscorrelation demerit factor is
\[
\CDF(f,g) = \cnrd{f}{g}
\]
and the autocorrelation demerit factor is
\begin{equation}\label{Penelope}
\ADF(f) = \frac{\normff{f}}{\normtf{f}}-1
\end{equation}
This links the work of Littlewood (see \cite{Littlewood-66} and \cite[Problem 19]{Littlewood-68}) on flatness of polynomials on the complex unit circle with the work of Golay \cite{Golay-72,Golay-75} on merit factors.

Sarwate \cite[eqs.~(13),(38)]{Sarwate} calculated expected values of demerit factors for randomly selected binary sequences (where each term is independent of the others and has equal probability of being $+1$ or $-1$).
For a randomly selected sequence $f$ of length $\ell$, Sarwate calculated the expected value of the autocorrelation demerit factor to be
\begin{equation}\label{Otto}
E[\ADF(f)] = 1 -\frac{1}{\ell}.
\end{equation}
For a randomly selected pair $(f,g)$ of sequences of length $\ell$, Sarwate calculated the expected value of the crosscorrelation demerit factor to be
\begin{equation}\label{Henry}
E[\CDF(f,g)] = 1.
\end{equation}
So typical values of both autocorrelation and crosscorrelation demerit factors will be about $1$ when the length $\ell$ is large, as it quite often is.
For example, Gold sequences of length $1023$ are used in the Global Positioning System (GPS), and code division multiple access communications (CMDA) protocols use even longer sequences.
Thus we want constructions that produce families of low correlation sequence pairs of various lengths.
Typically, our constructions produce families with unbounded lengths, and we rate a family by the {\it asymptotic demerit factors}, that is, the limit of the autocorrelation or crosscorrelation demerit factor as the length of the sequences tends to infinity.
For sequences derived from finite field characters, it has been observed in many cases \cite{Katz-16,Boothby-Katz} that the limiting behavior of families is approached quite rapidly, so that even sequences of quite modest length (of the order of a hundred or more) already have demerit factors close to the limiting values.

\section{High asymptotic autocorrelation merit factor}\label{Gary}

In this section we shall discuss constructions that give infinite families of binary sequences with high asymptotic autocorrelation merit factor.
Recall \eqref{Otto}, which says that randomly selected binary sequences of length $\ell$ have an average autocorrelation demerit factor of $1-1/\ell$, which is close to $1$ for large $\ell$.
It is possible to obtain families where the asymptotic demerit factor is considerably lower.
It is relatively rare to find such families, and to the author's best knowledge, the first one that was discovered derives from the Rudin-Shapiro polynomials, which shall be discussed further in Section \ref{Lawrence}.
It was Littlewood who originally proved a result tantamount to showing that this family of polynomials has asymptotic demerit factor $1/3$ \cite[pp.~27--28]{Littlewood-68}.
At the time, the concept of merit factor for correlation had not yet been defined: the formula for the autocorrelation merit factor would appear as a ``factor'' in a 1972 paper by Golay \cite{Golay-72}, who later called this the ``merit factor'' in another paper a few years later \cite{Golay-75}.
What Littlewood actually proved \cite[p.~28]{Littlewood-68} is a formula for the ratio of $L^4$ norm to $L^2$ norm of the Rudin-Shapiro polynomials, which via \eqref{Penelope} is equivalent to finding the asymptotic autocorrelation demerit factor.
The Rudin-Shapiro sequence family has one sequence $f_n$ of length $2^n$ for each nonnegative integer $n$, and Littlewood's result shows that $\ADF(f_n)=(1-(-1/2)^n)/3$, which tends to $1/3$ in the limit as $n$ tends to infinity.

If we consider Littlewood's result as the first low asymptotic demerit factor record, then this record was broken by H\o holdt and Jensen \cite{Hoholdt-Jensen} with cyclically shifted Legendre sequences, and that record was again broken by Jedwab, Katz, and Schmidt \cite[Theorem 1.1]{Jedwab-Katz-Schmidt-polynomials} with Legendre sequences that are cyclically shifted and appended (periodically extended).
The Legendre sequences and their modifications shall be discussed in more detail in Section \ref{Ursula}.
We summarize these records in Table \ref{Quentin}, which in addition to listing the asymptotic demerit factor also lists its reciprocal, the asymptotic merit factor, which is the way these results are usually presented in the literature.
The asymptotic demerit factor of $0.157\ldots$ listed for the shifted and appended Legendre sequences is the smallest real root of the polynomial $27 x^3 - 417 x^2 + 249 x - 29$.
\begin{table}[!ht]
\caption{Records for high asymptotic autocorrelation merit factor}\label{Quentin}
\begin{center}
{\setlength{\extrarowheight}{3pt}    
\begin{tabular}{l|c|c|l}
 & \multicolumn{2}{c|}{Asymptotic} & \finishtablerow
Sequence family & $\AMF$ & $\ADF$ & Proved by  \finishtablerow
\hline
\colortablerow
Rudin-Shapiro & $3$ & $0.333\ldots$ &  Littlewood (1968) \cite[pp.~28]{Littlewood-68} \finishtablerow
Legendre, shifted & $6$ & $0.166\ldots$ &  H\o holdt-Jensen (1988) \cite{Hoholdt-Jensen} \finishtablerow
\colortablerow
Legendre, shifted & & & Jedwab-Katz-Schmidt\finishtablerow
\colortablerow
\quad and appended  & \multirow{-2}{*}{$6.342\ldots$} & \multirow{-2}{*}{$0.157\ldots$} & (2013) \cite[Theorem 1.1]{Jedwab-Katz-Schmidt-polynomials} 
\end{tabular}
}
\end{center}
\end{table}
It should also be noted that, in addition to the records on Table \ref{Quentin}, an important advance was the determination by Jensen and H\o holdt \cite[\S 5]{Jensen-Hoholdt} of the asymptotic merit factor of a class of sequences known as maximal linear recursive sequences (m-sequences).
These sequences shall be described in the next section, but they are of great interest because it is easy to generate large families of them for use in communications networks.
Jensen and H\o holdt showed that any infinite family of m-sequences has asymptotic autocorrelation demerit factor $1/3$, which equals the performance of the Rudin-Shapiro polynomials.

\section{Sequences from additive characters}\label{Theresa}

In this section, we shall discuss sequences derived from additive characters of finite fields, of which the most fundamental are the {\it maximum length linear recursive shift register sequences}, which are also called {\it maximal linear recursive sequences}, or just {\it m-sequences}.
Let $\Fq$ be a finite field of characteristic $p$ and order $q=p^n$.
An {\it additive character} is a homomorphism from the additive group $\Fq$ to the multiplicative group $\Cu$.
We use $\Tr\colon \Fq\to\Fp$ to denote the absolute trace from $\Fq$ to its prime field $\Fp$.
Then for each $a\in\Fq$, the map $\epsilon_a\colon \Fq\to\Cu$ with $\epsilon_a(x)=\exp(2\pi i \Tr(a x)/p)$ is an additive character of $\Fq$, and $\{\epsilon_a : a \in\Fq\}$ is the entire group of $q$ additive characters from $\Fq$ into $\Cu$, with $\epsilon_0$ being the trivial character (that maps every element of $\Fq$ to $1$), while $\epsilon_1$ is called the {\it canonical additive character}.

Let $\alpha$ be a primitive element  of $\Fq$.
Let us list the nonzero elements of $\Fq$ as powers of the primitive element $\alpha$, that is, as
\[
\alpha^0,\alpha^1,\alpha^2,\ldots,\alpha^{q-2},
\]
and then apply a nontrivial additive character $\psi$ to obtain a sequence
\begin{equation}\label{Richard}
\left(\psi(\alpha^0),\psi(\alpha^1),\ldots,\psi(\alpha^{q-2})\right).
\end{equation}
An {\it m-sequence} is any sequence obtained in this way.
Any nontrivial additive character of a finite field of characteristic $p$ has the complex $p$th roots of unity as its outputs, so the m-sequences produced from fields of characteristic $p$ are $p$-ary sequences.
We shall mainly be interested in binary m-sequences, which derive from fields of characteristic $2$.

Changing the nontrivial character $\psi$ in \eqref{Richard} simply causes a cyclic shift of the m-sequence, and each of the $q-1$ nontrivial additive characters of $\Fq$ produces a different cyclic shift, and the $q-1$ different cyclically shifted versions of our m-sequence are all distinct.
If the character $\psi$ we use is the canonical additive character, we call the m-sequence produced in \eqref{Richard} a {\it Galois sequence} or a {\it naturally shifted m-sequence}.

Changing the primitive element $\alpha$ in \eqref{Richard} to another primitive element $\beta=\alpha^d$ (where $\gcd(d,q-1)=1$ to maintain primitivity) causes the m-sequence \eqref{Richard} to be decimated by $d$, that is, the new m-sequence based on $\beta$ is what one obtains by selecting every $d$th element from the original sequence (starting at the beginning and proceeding cyclically modulo the length $q-1$).
If $\beta$ is a Galois conjugate of $\alpha$ over the prime field $\Fp$, that is, if $d$ is a power of $p$ modulo $q-1$, then one just gets back the original sequence (up to some cyclic shift); otherwise one gets a sequence that is distinct from every cyclic shift of the original sequence.
Thus decimations $d$ that are a power of $p$ modulo $q-1$ are said to be {\it degenerate}.
When the original sequence is a Galois sequence, decimation by a degenerate $d$ yields back the original sequence exactly (not even cyclically shifted).
Another type of decimation that will become useful later in Section \ref{Orestes} is a {\it reversing decimation}, which is any decimation $d$ for which there is an integer $k$ such that $d \equiv -p^k \pmod{q-1}$.
If we decimate an m-sequence by such a $d$, one obtains the reverse of the original sequence (up to a cyclic shift).

We can now count the total number of m-sequences, based on our freedom to choose the character (cyclic shifting) and the primitive element modulo Galois conjugacy (decimation).
If we organize m-sequences of length $p^n-1$ into classes of sequences modulo cyclic shifting (with $p^n-1$ sequences per class), the number of classes of m-sequences will be equal to the number of classes of primitive elements of $\Fq=\Fpn$ modulo Galois conjugacy (with $n$ Galois conjugates per class).
Since the number of primitive elements in $\Fq=\Fpn$ is $\phi(p^n-1)$, where $\phi$ is Euler's $\phi$-function, the total number of m-sequences of length $p^n$ is $(p^n-1)\phi(p^n-1)/n$.
If $\alpha$ is a primitive element of our field $\Fq$, then $\alpha^{-1}$ will also be a primitive element and will not be a Galois conjugate of $\alpha$ unless $q \leq 4$ (in which case all primitive elements in $\Fq$ are Galois conjugates of each other).
So $\phi(p^n-1)/n > 1$ whenever $p^n > 4$, in which case it is possible to obtain at least two cyclically distinct m-sequences (related by a nondegenerate decimation) of length $p^n-1$.

Our $p$-ary m-sequence \eqref{Richard} of length $p^n-1$ follows a linear recursion of depth $n$ whose characteristic polynomial is the minimal polynomial of the primitive element $\alpha$ over the prime field $\Fp$.
Because of this, our m-sequence of length $p^n-1$ can be generated using a linear feedback shift register of length $n$.
This efficient generation of very long sequences with rather small circuits makes m-sequences very popular in applications.
Furthermore, from two m-sequences of length $p^n-1$ related by a nondegenerate decimation $d$, one can construct a family of $p^n+1$ Gold sequences of length $p^n-1$.
Gold's original construction \cite[\S IV]{Gold} uses carefully chosen decimations $d$ to produce families where all the sequences have low periodic autocorrelation and all the pairs have low periodic crosscorrelation.

We now give an overview of findings on the asymptotic aperiodic autocorrelation merit factor of binary m-sequences and their relatives.
As mentioned in the previous section, Jensen and H\o holdt \cite[\S 5]{Jensen-Hoholdt} proved that m-sequences have asymptotic autocorrelation demerit factor $1/3$.
Jedwab and Schmidt \cite[Theorems 11 and 12]{Jedwab-Schmidt-10} applied some constructions described by Parker \cite[Lemmas 3 and 4]{Parker} to m-sequences to produce families of related sequences that also have asymptotic autocorrelation demerit factor $1/3$.
Parker gave two constructions, each of which takes a sequence as an input and gives a longer sequence as an output.
The first construction, called the {\it negaperiodic construction}, doubles the length of the sequence, so we shall call it {\it Parker's doubling construction}.
The second construction, called the {\it periodic construction}, quadruples the length of the sequence, so we shall call it {\it Parker's quadrupling construction}.

Another technique that was used to modify m-sequences is {\it appending}, which originates with studies by Kirilusha and Narayanaswamy \cite{Kirilusha-Narayanaswamy}.
We let $f(z)=\sum_{j=0}^{\ell-1} f_j z^j$ be a sequence of length $\ell$ (represented in polynomial form), and extend the definition of $f_j$ so that $f_{j+\ell}=f_j$ for all $j \in \Z$.
Then we can truncate or periodically extend $f$ simply by changing the range of summation.
For example if $m < \ell$, then $g(z)=\sum_{j=0}^{m-1} f_j z^j$ is a truncated version of $f$, while if $m > \ell$, then it is an periodically extended version of $f$.
In these respective cases, we say that this new sequence $g$ is {\it $f$ truncated to $m/\ell$ times its usual length} or {\it $f$ appended to $m/\ell$ times its usual length}.
Jedwab, Katz, and Schmidt \cite[Theorem 2.2]{Jedwab-Katz-Schmidt-advances} proved that if one applies this procedure to m-sequences, one can produce families with an asymptotic autocorrelation demerit factor of $0.299\ldots$, which is the smallest real root of the polynomial $3 x^3 - 33 x^2 + 33 x - 7$.
To achieve this, one should use m-sequences appended to about $1.115\ldots$ times their usual length, where $1.115\ldots$ is the middle root of $x^3-12 x+12$.
Jedwab, Katz, and Schmidt also combined the appending procedure with Parker's constructions to produce further families with asymptotic autocorrelation demerit factor $0.299\ldots$.

The concept of an m-sequence can be generalized to a produce a larger family of sequences called the {\it Gordon-Mills-Welch sequences} \cite{Scholtz-Welch}.
The construction of Gordon-Mills-Welch sequences differs from that of m-sequences in that the character $\psi$ used in the m-sequence construction \eqref{Richard} is replaced with a ``twisted'' version.
G\"unther and Schmidt \cite[p.~344--345]{Guenther-Schmidt} have recently shown Gordon-Mills-Welch sequences attain the same asymptotic autocorrelation demerit factors that m-sequences do: $1/3$ for natural length and $0.299\ldots$ if appended.

\section{Sequences from multiplicative characters}\label{Ursula}

Now we describe a sequence construction that is in some sense dual to the construction of m-sequences.
The pseudorandom behavior of m-sequences can be traced to the fact that we form them by listing the nonzero elements of a finite field $\Fq$ in an order based on the {\it multiplicative structure} of the field (that is, as powers of a primitive element) and then apply an {\it additive character} to them (see \eqref{Richard} and the commentary preceding it).
Our next construction is dual in the sense that we shall devise a listing of the elements of a finite field based on the {\it additive structure} of the field and then apply a {\it multiplicative character} to them.

To make a multiplicative character sequence, let us start with a finite field $\Fp$ of prime order $p$, and write its elements in an order based on the additive structure of the field.
We can take $1$ as our additive generator, and then the additive analogue of taking increasing powers of this element is to form sums of increasing numbers of $1$, that is, we list the elements of $\Fp$ in the order
\begin{equation}\label{Albert}
0,1,1+1=2,1+1+1=3,\ldots,p-1.
\end{equation}
Now let $\chi \colon \Fp \to \Cu$ be a multiplicative character, that is, a group homomorphism from $\Fpu$ to $\Cu$, and make sure that $\chi$ is nontrivial, that is, does not map every element to $1$.
Normally one extends a multiplicative character $\chi$ by setting $\chi(0)=0$.
Then we apply our nontrivial multiplicative character $\chi$ to our list \eqref{Albert} of elements of $\Fp$ to obtain the sequence
\begin{equation}\label{Zachary}
\left(\chi(0),\chi(1),\chi(2),\ldots,\chi(p-1)\right).\footnote{This sequence was formed using the specific choice of $1$ as the additive generator of $\Fp$.  We could have replaced $1$ with any other $a\in\Fpu$ to form the list $0,a,2a,\ldots,(p-1)a$ of elements of $\Fp$ instead of \eqref{Albert}, and then apply $\chi$ to every term to get $\left(\chi(0),\chi(a),\chi(2 a),\ldots,\chi((p-1)a)\right)$.  This would just give the sequence in \eqref{Zachary} multiplied by the unimodular scalar $\chi(a)$.  This scalar multiplciation has no effect on the magnitudes of correlation values.}
\end{equation}
The multiplicative characters of $\Fp$ form a cyclic group of order $p-1$ under multiplication.
If $\chi$ is a character whose order is $m$ in this group, then all terms except $\chi(0)=0$ are $m$th roots of unity.
Normally, we replace the initial $\chi(0)=0$ term with a complex number $m$th root of unity (typically one just uses $1$) to get a true $m$-ary sequence.

If $p$ is an odd prime, then the group of multiplicative characters of $\Fp$ always contains one and only one character of order $2$, which is called the {\it quadratic character} or {\it Legendre symbol}.
If we use this as our character $\chi$ in the construction above (and replace $\chi(0)$ with $1$), then we obtain a binary sequence $h=(h_0,h_1,\ldots,h_{p-1})$,  called a {\it Legendre sequence}, where
\begin{equation}\label{Veronica}
\begin{split}
h_j = \begin{cases}
+1 & \text{if $j$ is the square of some element in $\Fp$}, \\
-1 & \text{if $j$ is not the square of any element in $\Fp$}.
\end{cases}
\end{split}
\end{equation}
Since there is only one character of order $2$ over each prime field $\Fp$ of odd order, this construction gives us only one binary sequence of length $p$ for each odd prime $p$.
Contrast this with the construction of m-sequences in Section \ref{Theresa}, which often produces many sequences of the same length that are not related to each other by cyclic shifting.

One might ask why we only used prime fields in the construction of multiplicative character sequences, while we used arbitrary finite fields to construct m-sequences.
The reason is that prime fields are the only finite fields that are cyclic groups under addition, so they are the only finite fields where one can generate a list of all the elements using a single additive generator.
A finite field $\Fpn$ of characteristic $p$ and order $p^n$ is an $n$-dimensional vector space over $\Fp$, so $\Fpn$ can be generated by an $\Fp$-basis consisting of $n$ elements.
Using this $n$-dimensional description of $\Fpn$, we can generalize our construction to create $n$-dimensional arrays whose entries are given by evaluations of multiplicative characters, and there are natural definitions of correlation for these arrays, with many results analogous to what we present about sequences in this paper, for example, see \cite{Katz-13}.

As noted above, the standard multiplicative character construction only gives one binary sequence of length $p$ for each odd prime $p$.
This is not very satisfactory if we are interested in finding pairs or larger families of binary sequences with low crosscorrelation.
Boothby and Katz \cite{Boothby-Katz} discovered that one can often obtain sequences with good aperiodic autocorrelation and crosscorrelation properties using linear combinations of multiplicative characters.
Among the sequences formed from linear combinations of multiplicative characters that Boothby and Katz studied are the {\it cyclotomic sequences}, whose periodic and aperiodic autocorrelation properties had been studied by Boehmer \cite{Boehmer}, and whose periodic autocorrelation and periodic crosscorrelation properties had been studied by Ding, Helleseth, and Lam \cite{Ding-Helleseth-Lam-99,Ding-Helleseth-Lam-00}.

We now describe the construction of cyclotomic sequences.
Let $m$ be an even positive integer and let $p$ be a prime with $m\mid p-1$.
Then let $\Fpum$ be the set $\{a^m: a\in \Fpu\}$ of $m$th powers, which is a subgroup of order $(p-1)/m$ in the group in $\Fpu$.
We form the quotient group $\cycquot$ of order $m$, which consists of $m$ cosets of $\Fpum$ in $\Fpu$.
Partition $\Fp$ into two sets, $A$ and $B$ as follows: $A$ contains $0$ along with the union of $m/2$ cosets of $\Fpum$, while $B$ contains the union of the other $m/2$ cosets of $\Fpum$.
Then we define a sequence $f=(f_0,f_1,\ldots,f_{p-1})$ where $f_j=1$ if $j \in A$ and $f_j=-1$ if $j \in B$.
The choices that we make when allocating cosets to $A$ or $B$ can influence the correlation behavior of the sequences.

Let us consider cyclotomic sequences in some of the simplest cases , that is, when $m$ is small.
When $m=2$ and when we define $A$ to be $\{0\} \cup \Fputwo$, we recover the Legendre sequence $h$ defined above (cf.~\eqref{Veronica}).
When $m=4$, we define two new sequences in this manner: let $\alpha$ be a primitive element of $\Fp$ and list the four cosets of $\Fpufour$ as $R_j=\alpha^j \Fpufour$ for $j \in \{0,1,2,3\}$.
Then define $f=(f_0,f_1,\ldots,f_{p-1})$ by
\begin{equation}\label{William}
\begin{split}
f_j = \begin{cases}
+1 & \text{if $j \in R_0 \cup R_1 \cup\{0\}$} \\
-1 & \text{if $j \in R_2 \cup R_3$},
\end{cases}
\end{split}
\end{equation}
and define $g=(g_0,g_1,\ldots,g_{p-1})$ by
\begin{equation}\label{Xavier}
\begin{split}
g_j = \begin{cases}
+1 & \text{if $j \in R_0 \cup R_3 \cup\{0\}$} \\
-1 & \text{if $j \in R_1 \cup R_2$}.
\end{cases}
\end{split}
\end{equation}
The Legendre sequence defined in \eqref{Veronica} reappears in this formalism as $h=(h_0,h_1,\ldots,h_{p-1})$, where
\begin{equation}\label{Yolanda}
\begin{split}
h_j = \begin{cases}
+1 & \text{if $j \in R_0 \cup R_2 \cup\{0\}$} \\
-1 & \text{if $j \in R_1 \cup R_3$},
\end{cases}
\end{split}
\end{equation}
because $R_0 \cup R_2 = \Fputwo$.

As mentioned above, these sequences are all formed by applying linear combinations of multiplicative characters to the list \eqref{Albert}.
In the case of our Legendre sequence $h$, the linear combination is just the single quadratic character.
For sequences $f$ and $g$, we let $\theta$ be the multiplicative character of $\Fp$ of order $4$ defined by $\theta(\alpha^j)=e^{\pi i j/2}=i^j$, where we recall that $\alpha$ is the primitive element of $\Fp$ used to define the cosets $R_j$.
The other multiplicative character of $\Fp$ of order $4$ is $\conj{\theta}$, which has $\conj{\theta}(\alpha^j)=(-i)^j$.
To get the sequence $f$, one first applies the linear combination of characters
\[
\lambda(x)=\frac{1-i}{2} \theta(x) + \frac{1+i}{2} \conj{\theta}(x)
\]
to the elements of the list \eqref{Albert} to get the sequence
\[
\left(\lambda(0)=0,\lambda(1),\ldots,\lambda(p-1)\right),
\]
and then one replaces $\lambda(0)=0$ with $1$.
To get the sequence $g$, one uses the same procedure, but with 
\[
\mu(x)=\frac{1+i}{2} \theta(x) + \frac{1-i}{2} \conj{\theta}(x)
\]
in place of $\lambda$.

Now that we have introduced our sequences, we need to discuss the modified versions of them that have been found to have good correlation properties.
First of all, unlike the definition of m-sequences in Section \ref{Theresa}, the above definitions of sequences derived from multiplicative characters do not embrace all cyclic shifts of a given sequence.
Instead each sequence comes defined with a particular cyclic shift.
We will want to cyclically shift our multiplicative character sequences, but then we call them {\it shifted multiplicative character sequences} to distinguish them from the originals.
For example, Golay \cite{Golay-83} reported Turyn's discovery that one can significantly increase the autocorrelation merit factor of Legendre sequences if one cyclically shifts them.
We can also apply Parker's doubling and quadrupling constructions, as well as the appending technique described in Section \ref{Theresa} to sequences produced from the constructions mentioned above.

As mentioned in Section \ref{Gary}, H\o holdt and Jensen \cite{Hoholdt-Jensen} proved that appropriately cyclically shifted Legendre sequences achieve asymptotic autocorrelation demerit factor $1/6$.
One can also obtain the same asymptotic demerit factor with sequences formed from Legendre sequences with Parker's doubling construction, as proved by Xiong and Hall \cite[Theorem 3.3]{Xiong-Hall-08}, or with Parker's quadrupling construction (combined with appropriate shifting), as proved by Schmidt, Jedwab, and Parker \cite[Theorem 8]{Schmidt-Jedwab-Parker}.

If one also allows appending (with or without Parker's constructions and with appropriate shifting) of Legendre sequences, then Jedwab, Katz, and Schmidt (in \cite[Theorem 1.1]{Jedwab-Katz-Schmidt-polynomials} an \cite[Theorem 2.1]{Jedwab-Katz-Schmidt-advances}) showed that one can achieve an asymptotic autocorrelation demerit factor of $0.157\ldots$, which is the smallest real root of the polynomial $27 x^3 - 417 x^2 + 249 x - 29$.
To achieve this, one appends the sequences to be about $1.057\ldots$ times their usual length, where $1.057\ldots$ is the middle root of $4 x^3-30 x+27$.

We note that one can generalize the notion of Legendre sequences, which are based on quadratic characters of prime fields, to {\it Jacobi sequences}, which are based on quadratic characters of integer residue rings (which allows for composite lengths).
Jacobi sequences and their modifications (using shifting, Parker's constructions, and appending) often behave similarly to Legendre sequences, and are able to achieve the same asymptotic autocorrelation demerit factor of $1/6$ in their natural lengths and $0.157\ldots$ when appended.
Although there are some detailed conditions that must be respected to obtain this behavior, Jacobi sequences provide good autocorrelation at a wider variety of lengths than one could obtain with Legendre sequences alone.
See the papers of Jensen, Jensen, and H\o holdt \cite[Theorem 2.4, \S\S IV--V]{Jensen-Jensen-Hoholdt}; Xiong and Hall \cite[\S V]{Xiong-Hall-08}, \cite{Xiong-Hall-11}; Jedwab and Schmidt \cite{Jedwab-Schmidt-12}; and Jedwab, Katz, and Schmidt \cite[Theorem 2.3, Corollary 2.4, and \S 6]{Jedwab-Katz-Schmidt-advances} for the principal results.

Boothby and Katz \cite[Theorem 19]{Boothby-Katz} and G\"unther and Schmidt \cite[p.~347]{Guenther-Schmidt} show that carefully selected families of the cyclotomic sequences can produce asymptotic demerit factor $1/6$ when suitably cyclically shifted, and if one also appends appropriately, this can be improved to the same $0.157\ldots$ value as for appended Legendre sequences.
When designing a particular family of such sequences it is necessary to make judicious choices of which cyclotomic classes to assign $+1$ values and which cyclotomic classes to assign $-1$ values, otherwise the demerit factors will be bounded away from $0.157\ldots$.
Boothby and Katz \cite[Theorem 10]{Boothby-Katz} give conditions under which one can achieve limiting autocorrelation demerit factor $0.157\ldots$ for sequences derived from linear combinations of multiplicative characters, which includes the binary cyclotomic sequences as a proper subclass.
Boothby and Katz's conditions are met by the cyclotomic sequences derived from quartic characters described at \eqref{William} and \eqref{Xavier}, and these sequences were used by both Boothby and Katz \cite[Theorem 19]{Boothby-Katz} and G\"unther and Schmidt \cite[p.~347]{Guenther-Schmidt} as examples showing that one can achieve asymptotic demerit factor $0.157\ldots$.
G\"unther and Schmidt also give examples of sequence designs from six cyclotomic classes, and exhibit families that achieve limiting demerit factor $0.157\ldots$ and families that do not.

Even if one uses sequence designs from four or six cyclotomic classes that can achieve asymptotic autocorrelation demerit factor $0.157\ldots$, one must restrict one's sequence family to contain only sequences derived from fields $\Fp$ whose orders fulfill exacting number-theoretic conditions (see \cite[Theorem 19]{Boothby-Katz} and \cite[Corollaries 2.5 and 2.6]{Guenther-Schmidt}).
As such, if one adopts one of these cyclotomic sequence designs, the sequences produced will, at most lengths, fall short of Legendre sequences in terms of autocorrelation performance.
Accordingly Boothby and Katz \cite[p.~6162]{Boothby-Katz} point out that there is little reason to use these cyclotomic sequences in applications where one wants a single sequence with good autocorrelation performance; rather, the real interest of cyclotomic sequences is that there is more than one of them of a given length, so they can be used in applications where crosscorrelation is important.
We shall discuss this further in Section \ref{Orestes}.

G\"unther and Schmidt \cite[p.~344--345]{Guenther-Schmidt} also studied another family of pseudorandom sequences called the {\it Sidel\,{\cprime}\!nikov sequences} \cite{Sidelnikov}.
These sequences are derived from quadratic characters of finite fields, but in a different way than Legendre sequences.
G\"unther and Schmidt proved that Sidel{\cprime}nikov sequences have the same asymptotic autocorrelation demerit factors as m-sequences: $1/3$ in their natural length, and $0.299\ldots$ for appropriately appended versions.

\section{Rudin-Shapiro-like sequences}\label{Lawrence}

In his master's thesis \cite[p.~42]{Shapiro}, Shapiro devised a construction of a family $f_0,f_1,f_2,\ldots$ of sequences, where $f_n$ is a binary sequence of length $2^n$.
Shapiro's construction is easier to understand when one introduces a companion family of sequences, $g_0,g_1,g_2,\ldots$.
Recall from Section \ref{Katherine} our identification of sequences with polynomials.
The construction is the recursion with
\begin{align}\label{Cecilia}
\begin{split}
f_0(z)  & = g_0(z) =1 \\
f_{n+1}(z) & = f_n(z) + z^{2^n} g_n(z) \\
g_{n+1}(z) & = f_n(z) - z^{2^n} g_n(z).
\end{split}
\end{align}
In terms of sequences, this says that $f_{n+1}$ is the concatenation of $f_n$ and $g_n$, while $g_{n+1}$ is the concatenation of $f_n$ and $-g_n$.
Shapiro's sequences (polynomials) are what one gets when one retains $f_0,f_1,f_2,\ldots$ and discards the companion sequences.
These sequences were rediscovered by Rudin \cite[eq.~(1.5)]{Rudin} somewhat later, and are now known as {\it Rudin-Shapiro sequences} (or {\it Rudin-Shapiro polynomials}).

Around the same time as Shapiro, Golay \cite{Golay-51} discovered an equivalent construction that produced what he called {\it complementary pairs} (now called {\it Golay complementary pairs} or just {\it Golay pairs}).
These are pairs $(f,g)$ of sequences of the same length with $C_{f,f}(s)+C_{g,g}(s)=0$ for all $s\not=0$, and Golay originally devised them for use in multislit spectrometry.
We say that a Golay pair has length $\ell$ to mean that it consists of two sequences, each of length $\ell$.
If one pairs the Shapiro sequences with their companions, that is, if one considers $(f_0,g_0), (f_1,g_1), (f_2,g_2), \ldots$, then one obtains one infinite family of complementary pairs constructed by Golay.
As mentioned in Section \ref{Gary} above, Littlewood \cite[pp.~27--28]{Littlewood-68} performed a calculation tantamount to showing that $\ADF(f_n)=(1-(-1/2)^n)/3$, which proves that the family $f_0,f_1,f_2,\ldots$ of Rudin-Shapiro sequences has asymptotic demerit factor $1/3$.

Brillhart and Carlitz \cite[Theorem 1]{Brillhart-Carlitz} showed that the companion sequences in the construction \eqref{Cecilia} were related to the main sequences by
\[
g_n(z)=(-1)^{n} z^{2^n-1} f_n(-1/z)
\]
for every $n$.
For any polynomial $h(z) \in \C[z]$, we define the {\it reciprocal polynomial of $h(z)$}, denoted $h^*(z)$, to be $z^{\deg h} h(1/z)$, that is, the polynomial obtained from $h$ by writing the coefficients in reverse order.
Then the result of Brillhart and Carlitz becomes $g_n(z)=(-1)^{2^n+n-1} f_n^*(-z)$, so that we could restate the construction without the companion sequences:
\begin{align}\label{David}
\begin{split}
f_0(z) & = 1 \\
f_{n+1}(z) & = f_n(z)+ (-1)^{2^n+n-1} z^{\deg{f_n+1}} f_n^*(-z).
\end{split}
\end{align}
It turns out that one gets similar asymptotic autocorrelation behavior no matter how the sign is chosen on the second term, as observed by H\o holdt, Jensen, and Justesen \cite[Theorem 2.3]{Hoholdt-Jensen-Justesen}, so we may generalize construction \eqref{David} to
\begin{align}\label{Esther}
\begin{split}
f_0(z) & = 1 \\
f_{n+1}(z) & = f_n(z)+ \sigma_n z^{\deg{f_n+1}} f_n^*(-z),
\end{split}
\end{align}
where $\sigma_0,\sigma_1,\ldots$ is any sequence of values in $\{+1,-1\}$, called the {\it sign sequence} for our construction.
H\o holdt, Jensen, and Justesen show \cite[Theorem 2.3]{Hoholdt-Jensen-Justesen} that regardless of the choice of sign sequence, one still obtains $\ADF(f_n)=(1-(-1/2)^n)/3$, so the asymptotic autocorrelation demerit factor is $1/3$.

Construction \eqref{Esther} was further generalized by Borwein and Mossinghoff \cite[pp.~1159 and 1161]{Borwein-Mossinghoff}, by allowing much more freedom at the start:
\begin{align}\label{Fiona}
\begin{split}
f_0(z) & = \text{any polynomial with coefficients in $\{+1,-1\}$} \\
f_{n+1}(z) & = f_n(z)+ \sigma_n z^{\deg{f_n+1}} f_n^*(-z),
\end{split}
\end{align}
where again $\sigma_0,\sigma_1,\ldots$ is any sequence of values in $\{+1,-1\}$, called the {\it sign sequence} for our construction.
We call $f_0$ the {\it seed} of the construction, and we call the family $f_0,f_1,f_2,\ldots$ of polynomials the {\it stem} generated by that seed and sign sequence.
Following Borwein and Mossinghoff, we call families of sequences (polynomials) generated from construction \eqref{Fiona} {\it Rudin-Shapiro-like sequences (polynomials)}.

Borwein and Mossinghoff found a precise formula \cite[Theorem 1]{Borwein-Mossinghoff} for the autocorrelation demerit factor of Rudin-Shapiro-like polynomials produced by their construction \eqref{Fiona}.
That is, they have a formula for computing $\ADF(f_n)$ for every $n$, and from this they compute the asymptotic autocorrelation merit factor, which depends on the seed but not the sign sequence.
They show that the asymptotic autocorrelation demerit factor is always greater than or equal to $1/3$, but only achieves a value of $1/3$ for certain seeds, which we call {\it optimal seeds}.
Borwein and Mossinghoff performed a computer search (informed by some of their theoretical results) over all binary sequences of length $40$ or less, and found optimal seeds of lengths $1$, $2$, $4$, $8$, $16$, $20$, $32$, and $40$, and no optimal seeds of other lengths in this range.
Later Katz, Lee, and Trunov \cite[Table 1]{Katz-Lee-Trunov} performed a larger search that extended to all seeds of length $52$ or less, and found new optimal seeds at length $52$ (but at no other length between $40$ and $52$).
This data was explained by the following classification of optimal seeds \cite[Theorem 1]{Katz-Trunov}: a seed of length greater than $1$ is optimal if and only if it is the interleaving of a Golay complementary pair, where the {\it interleaving} of two sequences $a=(a_0,a_1,\ldots,a_{\ell-1})$ and $b=(b_0,b_1,\ldots,b_{\ell-1})$ of length $\ell$ is the the sequence $(a_0,b_0,a_1,b_1,\ldots,a_{\ell-1},b_{\ell-1})$ of length $2\ell$.
In polynomial terms, the interleaving is $a(z^2)+z b(z^2)$.
This result, along with the known fact that the two seeds ($+1$ and $-1$) of length $1$ are optimal, gives a full classification of the optimal seeds.
A construction of Turyn \cite[Corollary to Lemma 5]{Turyn} shows that there is a Golay complementary pair of length $2^a 10^b 26^c$ for every choice of nonnegative integers $a$, $b$, $c$.
Thus there are infinitely many optimal seeds.

\section{High asymptotic crosscorrelation merit factor}\label{Rebecca}

Consider the sequences
\begin{align*}
f_\ell & = (+1,+1,+1,+1,\ldots,+1,+1) \\
g_\ell & = (+1,-1,+1,-1,\ldots,+1,-1)
\end{align*}
of even length $\ell$.
It is not difficult to calculate that
\[
\CDF(f_\ell,g_\ell) =\frac{1}{\ell},
\]
so that the asymptotic crosscorrelation demerit factor of the family of pairs $\{(f_\ell,g_\ell): \ell \in 2\Z\}$ is zero (so asymptotic crosscorrelation merit factor is infinite).
But it is also not difficult to calculate that
\[
\ADF(f_\ell)=\ADF(g_\ell)=\frac{2\ell^2+1}{3 \ell},
\]
so that the families $\{f_\ell: \ell \in 2\Z\}$ and $\{g_\ell: \ell \in 2\Z\}$ both have infinite asymptotic autocorrelation demerit factor (so asymptotic autocorrelation merit factor is $0$).
Thus it is not interesting to seek families of sequence pairs with low asymptotic crosscorrelation demerit factor in isolation from the asymptotic autocorrelation demerit factor of the constituent sequences.
What we really want to know is whether there is a way to make asymptotic autocorrelation and crosscorrelation demerit factors small at the same time.
In the next section we explore a measure that will help us quantify this goal.

\section{Pursley-Sarwate Criterion}\label{Nina}

Pursley and Sarwate \cite[eqs.~(3),(4)]{Pursley-Sarwate} proved that any pair $(f,g)$ of binary sequences has
\begin{equation}\label{Ignatius}
1-\sqrt{\ADF(f) \ADF(g)} \leq \CDF(f,g) \leq 1+\sqrt{\ADF(f) \ADF(g)}.
\end{equation}
Their proof is based on the Cauchy-Schwarz inequality.
We define the {\it Pursley-Sarwate criterion} for a pair $(f,g)$ of sequences to be
\[
\PSC(f,g)=\sqrt{\ADF(f) \ADF(g)}+\CDF(f,g),
\]
and then the bound \eqref{Ignatius} tells us that
\begin{equation}\label{James}
\PSC(f,g) \geq 1.
\end{equation}
We would like sequence pairs $(f,g)$ with $\ADF(f)$, $\ADF(g)$, and $\CDF(f,g)$ as small as possible, but the bound \eqref{James} shows that we cannot make them all simultaneously close to zero.
In view of Sarwate's expected values of demerit factors for randomly selected binary sequences in \eqref{Otto} and \eqref{Henry}, we expect a typical randomly selected pair $(f,g)$ of sequences to have $\PSC(f,g)$ of about $2$.
We would like to construct sequence pairs $(f,g)$ with $\PSC(f,g)$ as close to $1$ as possible.
We often consider {\it asymptotic $\PSC$} of families of sequence pairs, that is, the limiting value of $\PSC$ as the length of the sequences tends to infinity.

\section{Pairs with low asymptotic Pursley-Sarwate criterion}\label{Orestes}

One should recall the sequence constructions described in Sections \ref{Theresa}, \ref{Ursula}, and \ref{Lawrence} above: we now consider pairs of such sequences that have low Pursley-Sarwate criterion.
Table \ref{Margaret} lists some constructions that produce families of binary sequence pairs with low asymptotic $\PSC$.

\begin{table}[!ht]
\caption{Families of sequence pairs with low asymptotic $\PSC$}\label{Margaret}
\begin{center}
{\setlength{\extrarowheight}{3pt}\begin{tabular}{lccc}
Sequence pair $(f,g)$ & \multicolumn{3}{c}{Asymptotic Values} \\
construction & $\ADF(f)=\ADF(g)$ & $\CDF(f,g)$ & $\PSC(f,g)$ \\
\hline
\multicolumn{4}{c}{Katz (2016) \cite[pp.~5240, 5247]{Katz-16}} \finishtablerow
m-sequences, typical & $1/3$ & $1$ & $4/3$ \finishtablerow
m-sequence, reversing & $1/3$ & $5/6$ & $7/6$ \finishtablerow
half Legendre & $7/12$ & $7/12$ & $7/6$ \finishtablerow
\hline
\colortablerow
\multicolumn{4}{c}{Boothby-Katz (2017) \cite[pp.~6160--6161]{Boothby-Katz}} \finishtablerow
\colortablerow
quartic cyclotomics & in $[1/6,5/6]$ & in $[1/3,1]$  & $7/6$ \finishtablerow
\colortablerow
Legendre + quartic  & $1/6$ & $1$ & $7/6$ \finishtablerow
\hline
\multicolumn{4}{c}{Katz-Lee-Trunov \cite[Table 3]{Katz-Lee-Trunov}} \finishtablerow
Rudin-Shapiro-like & $1/3$ & $77/100$ & $331/300$  \finishtablerow 
\hline
\colortablerow
\multicolumn{4}{c}{Katz-Moore \cite[Theorem 1.1]{Katz-Moore}} \finishtablerow
\colortablerow
Golay pair & $1/3$ & $2/3$ & $1$ \finishtablerow \hline
\end{tabular}}
\end{center}
\end{table}

Let us provide some context and details for the table entries.
If we fix a $d \in \Z$ with $|d|$ not a power of $2$ and produce an infinite family of binary m-sequence pairs $(f_n,g_n)$ where $g_n$ is (up to cyclic shift) a decimation of $f_n$ by $d$, then Katz \cite[Theorem 1]{Katz-16} showed that the asymptotic crosscorrelation demerit factor will tend to $1$.
Since we have seen in Section \ref{Theresa} that the autocorrelation demerit factor tends to $1/3$, this produces a sequence family with asymptotic $\PSC$ of $4/3$.
We call this a {\it typical m-sequence construction}.
If we instead use $d=-2^k$ for some nonnegative integer $k$, then we produce a family of binary m-sequence pairs $(f_n,g_n)$ where $g_n$ is related to $f_n$ by the reversing decimation, and then Katz \cite[Theorem 2]{Katz-16} showed that one can lower the asymptotic crosscorrelation demerit factor to $5/6$ by appropriately cyclically shifting the sequences.
This results in families with asymptotic $\PSC$ of $7/6$.
We call this a {\it reversing m-sequence construction}.
(We never allow $d$ to be a power of $2$, since that will give degenerate decimations, and we will be correlating an m-sequence with cyclic shifts of itself.)

Another construction of Katz \cite[p.~5247]{Katz-16} takes a Legendre sequence (which has length equal to some odd prime $p$), cyclically shifts it in a certain way, discards the last term, and cuts the remaining sequence into a pair of two sequences of length $(p-1)/2$.
In this way one can obtain a family of sequence pairs $(f_n,g_n)$ with asymptotic $\ADF(f_n)$, $\ADF(g_n)$, and $\CDF(f_n,g_n)$ all equal to $7/12$, and thus asymptotic $\PSC$ equal to $7/6$.
We call this the {\it half Legendre construction}.

Boothby and Katz \cite[Theorem 21]{Boothby-Katz} crosscorrelated the two cyclotomic sequences \eqref{William} and \eqref{Xavier} derived from quartic characters, and also the cyclically shifted versions of these two sequences.
As mentioned in Section \ref{Ursula}, one obtains very low asymptotic autocorrelation demerit factor only for certain lengths, depending on a number-theoretic criterion.
It turns out that the crosscorrelation demerit factor of our sequence pairs tends to decrease as their autocorrelation demerit factor increases.
In fact, for any real number $A$ with $1/6\leq A \leq 5/6$, there is an infinite family of pairs $(f_n,g_n)$ of these cyclically shifted cyclotomic sequences such that asymptotic $\ADF(f_n)$ and $\ADF(g_n)$ are $A$, asymptotic $\CDF(f_n,g_n)$ is $7/6-A$, and asymptotic $\PSC$ is $7/6$.

One can also crosscorrelate cyclically shifted Legendre sequences (see \eqref{Veronica}, or equivalently \eqref{Yolanda}) with cyclically shifted versions of either of our quartic cyclotomic sequences (see \eqref{William} or \eqref{Xavier}).
In this case Boothby and Katz \cite[Theorem 20]{Boothby-Katz} show that one always obtains asymptotic $\CDF$ of $1$, so one should choose the shifted Legendre sequences and shifted quartic cyclotomic sequences to have limiting $\ADF$ of $1/6$, and thus obtain limiting $\PSC$ of $7/6$.

It should be noted that all the constructions of Katz and Boothby-Katz discussed here employ sequences in their usual length, but their results allow for the possibility of truncating and appending the sequences.
They showed \cite[eq.~(7)]{Boothby-Katz} that modest appending can be used to produce families of sequence pairs with asymptotic $\PSC$ slightly lower than $7/6$.

Katz, Lee, and Trunov crosscorrelated pairs of Rudin-Shapiro-like polynomials \cite[Theorem 2.4]{Katz-Lee-Trunov}, by beginning with two different seeds, $f_0$ and $g_0$, and applying recursion \eqref{Fiona} to produce two stems $f_0,f_1,\ldots$ and $g_0,g_1,\ldots$ (using the same sign sequence in recursion \eqref{Fiona} to produce the two stems).
They derive a precise formula for $\CDF(f_n,g_n)$ for each $n$.
This reduces to Borwein and Mossinghoff's precise formula \cite[Theorem 1]{Borwein-Mossinghoff} for $\ADF(f_n)$ when we set $g_n=f_n$ and subtract $1$ (see \eqref{Paul}).
From this one can compute $\PSC(f_n,g_n)$ precisely and from this determine the limiting $\PSC$.
Katz, Lee, and Trunov \cite[Table 3]{Katz-Lee-Trunov} found a pair of seeds, each of length $40$, that yield stems with limiting $\ADF$ of $1/3$ and limiting $\CDF$ of $77/100$, for a limiting $\PSC$ of $331/300=1.10\overline{3}$.

Finally, Katz and Moore \cite[Theorem 1.1]{Katz-Moore} proved that a pair $(f,g)$ of binary sequences has $\PSC(f,g)=1$ if and only if $(f,g)$ is a Golay complementary pair.
In Section \ref{Lawrence}, we noted that there are known to be Golay pairs of lengths $2^a 10^b 26^c$ for all nonnegative integers $a$, $b$, and $c$, so we have infinitely many binary sequence pairs with $\PSC$ exactly equal to $1$.
Thus we obtain an asymptotic $\PSC$ of $1$ with the Golay pairs.
We note that if $(f,g)$ is a Golay pair, then $\ADF(f)=\ADF(g)$.
The Golay pairs $(f_n,g_n)$ produced by recursion \eqref{Cecilia} have Rudin-Shapiro sequences as the first sequence in each pair.
Thus they have asymptotic $\ADF(f_n)$ equal to $1/3$ by the result of Littlewood described in Sections \ref{Gary} and \ref{Lawrence}.
So asymptotic $\ADF(g_n)$ is also $1/3$ for these pairs, and thus asymptotic $\CDF(f_n,g_n)$ is $2/3$.

We note that the families of sequence pairs on Table \ref{Margaret} all have equal asymptotic autocorrelation demerit factors for the first and second elements of the pairs.
While this must be the case for Golay complementary pairs, is not always observed in other constructions with low asymptotic $\PSC$.
For example, Katz, Lee, and Trunov \cite[Table 2]{Katz-Lee-Trunov} exhibit constructions of families $(f_0,g_0),(f_1,g_1),\ldots$ of pairs of Rudin-Shapiro-like sequences with low asymptotic $\PSC(f_n,g_n)$ where the asymptotic $\ADF(f_n)$ is not equal to the asymptotic $\ADF(g_n)$.

\section{Open questions}\label{Konrad}

We present two open questions that arise naturally from the considerations above.
\begin{question}
What is the lowest asymptotic autocorrelation demerit factor for binary sequences?
\end{question}
Or equivalently, what is the highest asymptotic autocorrelation merit factor for binary sequences?
Littlewood \cite[pp.~28--29]{Littlewood-68} made a conjecture that there is a infinite family of binary sequences with autocorrelation demerit tending to zero, or equivalently, autocorrelation merit factor tending to infinity.
Golay, on the other hand, conjectured that autocorrelation merit factor is bounded, and he proposed \cite{Golay-77} that asymptotic merit factor can never exceed $2 e^2 = 14.77\ldots$.  Later \cite{Golay-82} he revised his proposed upper bound on asymptotic merit factor to a value of about $12.32$.

We saw in Section \ref{Rebecca} a construction of sequence pairs with asymptotic crosscorrelation demerit factor of zero, but at the expense of poor autocorrelation performance.
It would be interesting to know how low asymptotic crosscorrelation demerit factor can be made without having poor autocorrelation demerit factors.
\begin{question}
Among infinite families of binary sequence pairs $(f,g)$ such that $\ADF(f)$, $\ADF(g)$, and $\CDF(f,g)$ tend to limits as the length of the sequences tends to infinity, and such that the limiting values for $\ADF(f)$ and $\ADF(g)$ are not greater than $1$, what is the lowest possible limiting value for $\CDF(f,g)$?
\end{question}
In Section \ref{Orestes} we saw that the construction of Boothby-Katz \cite[p.~6160]{Boothby-Katz} involving pairs of cyclotomic sequences derived from quartic characters furnishes families of sequence pairs $(f,g)$ with asymptotic $\CDF(f,g)$ of $1/3$ and asymptotic $\ADF(f)$ and $\ADF(g)$ of $5/6$ (so limiting $\PSC(f,g)$ is $7/6$).
If one wants to get even lower asymptotic $\CDF$, then one can use appending.
One would use Theorems 19 and 21 of \cite{Boothby-Katz} with the following parameters: one would let $\gamma=\pi/2$, let $\Lambda=1.207\ldots$ be the middle root of $4 x^3-36 x^2+60 x-27$, and let $R=(1-2\Lambda)/4$.
This would produce a family of pairs of quartic cyclotomic sequences that are cyclically shifted and then appended to $\Lambda=1.207\ldots$ times their normal length.
The limiting $\ADF$ of these sequences is $1$ and the limiting $\CDF$ is $0.254\ldots$, the middle root of $729 x^3 + 981 x^2 - 1245 x + 241$.
Note that the $\PSC$ for this family is $1.254\ldots$, which is considerably worse than the $7/6$ that one obtains without appending, so a considerable sacrifice in autocorrelation performance is being made for this increase in crosscorrelation performance.
One should note that when one appends like this, there will be a rather large value in the autocorrelation spectrum when the appended portion on one copy of the sequence comes into alignment with the initial portion of the other copy.
The magnitude of this large autocorrelation value will be about equal to the length of the appended sequence times $1-1/\Lambda=0.172\ldots$.

This paper has confined itself to analyzing the autocorrelation and crosscorrelation demerit factors of sequence pairs.
In many applications one needs larger families of sequences with low mutual correlation, and it would be interesting to extend the concepts here to that more general setting.

\section*{Acknowledgement}

The author thanks Yakov Sapozhnikov for his careful reading of this paper and his helpful suggestions.

\end{document}